\documentclass[a4paper,twocolumn,10pt]{article}
\usepackage{eurosis}
\usepackage{url}
\usepackage{hyperref}
\usepackage{cleveref}
\usepackage[numbers]{natbib}
\usepackage{booktabs}
\usepackage{graphbox}
\usepackage{svg}
\usepackage[document]{ragged2e}
\usepackage{float}
\usepackage{enumitem}
\usepackage[affil-it]{authblk}

\usepackage[obeyFinal,textsize=footnotesize,disable]{todonotes}
\bibpunct[; ]{(}{)}{,}{a}{}{;}

\begin{document}

\title{Designing a mobile game to generate player data --- lessons learned}

\author{William Wallis, William Kavanagh, Alice Miller \& Tim Storer\\ 
Department of Computing Science,\\University of Glasgow, \\
Scotland \\
email: \href{mailto:w.wallis.1@research.gla.ac.uk}{w.wallis.1@research.gla.ac.uk}
}
% \author{William Wallis and William Kavanagh and Alice Miller and Tim Storer}

% \email{w.wallis.1@research.gla.ac.uk}

% \thanks{Corresponding author}

% \affiliation{School of Computing Science, University of Glasgow, Scotland, Email: w.wallis.1@research.gla.ac.uk}
\date{}

% This form is inspired by CCSE's experience report: https://dl.acm.org/doi/pdf/10.1145/3368563

\maketitle

\thispagestyle{empty}

\keywords{Design, Prototyping, Datasets, Mobile Game Design}

\begin{abstract}

% Whole thing 
User friendly tools have lowered the requirements of high-quality game design to the point where researchers without development experience can release their own games. 
% Problem
However, there is no established best-practice as few games have been produced for research purposes.
% Solution
Having developed a mobile game without the guidance of similar projects, we realised the need to share our experience so future researchers have a path to follow.
% Paper contents
Research into game balancing and system simulation required an experimental case study, which inspired the creation of \emph{``RPGLite''}, a multiplayer mobile game.
% Summary
In creating RPGLite with no development expertise we learned a series of lessons about effective amateur game development for research purposes. In this paper we reflect on the entire development process and present these lessons. 

\end{abstract}

% =================================
\section{INTRODUCTION}

% General Problem
Procuring datasets to validate theoretical research findings can be difficult. Industrial sources for this data rarely have aligned interests with the researchers who require them. Academic datasets can often be over-specialised to the domain of the team who originally released them. While there is a requirement for more generally applicable and well-documented datasets for academic use, simple modern tooling has allowed for researchers to develop their own systems to obtain well-scoped datasets under their own control. 
\par

% What did we do?
Having developed techniques for generating synthetic gameplay data we required an appropriate dataset for comparison, so developed a mobile game to create it for us. This provided the opportunity for us to design a high-quality game to generate the data we required cheaply and easily. Since its release in April 2020, RPGLite\footnote{RPGLite is available from \href{https://rpglite.app/}{https://rpglite.app/}} has been more successful than we anticipated, populating a substantial dataset which will be of great value to our own research, and, we believe, to the wider research community. 
\par

% What are they about to read? % Why write an experience report
In this paper we detail our experience of creating RPGLite, including its planning, implementation, testing and deployment. The motivation for sharing this experience report was our own frustration at having nothing similar to support us when we embarked on this project. In providing reflections upon the successes and failures of our approach, we hope that this paper can guide other researchers considering a similar project.
\par

% key contributions of this work
The key contributions of this paper include: 
(i) Descriptions of the lessons learned in developing a mobile game for research purposes;
(ii) An outline of how a similar application can be released with no funding or development experience, and;
(iii) A frank discussion of the mistakes made and an analysis of how we could have produced a richer source of data more efficiently.
\par

% Table (/prose) of Contents
In the following section we describe why we needed to generate this dataset rather than use data from an already existing game. We go on to discuss the design of the game and the resulting implementation. In the body of the paper we present the four key lessons learned from the experience: how we came to learn them and how future researchers can use them to assist their development processes. Finally we summarise our contribution and detail future work which will follow as a result of our data collection from RPGLite.

% =================================
\section{MOTIVATION}\label{motivation}

% Explain the role of MCing in the previous work
In recent research we have developed an approach which uses model checking to analyse the balance and metagame development of a game. We refer to this approach as Chained Strategy Generation (CSG)~\citep{kavanagh2019balancing}. We use the PRISM model checker \citep{prism} and the PRISM-Games extension \citep{prism-G}, a probabilistic engine for analysis of various Markov models, including Discrete Time Markov Chains (DTMCs), which are purely probabilistic, and Markov Decision Processes (MDPs) which also involve non-deterministic choice. PRISM allows us to specify quantitative properties such as ``what is the probability that event $e$ happens?'' (for DTMCs), or ``for all possible sequences of choices, what is the greatest probability that event $e$ happens?'' (for MDPs). In our approach we define a model representing our game and use PRISM to determine player strategies of interest (here a strategy corresponds to a sequence of choices). For example, to determine a strategy for player 1 that corresponds to the best probability of winning, we check the property ``what is the maximum likelihood that player 1 wins the game?''. As well as returning this maximum probability, the model checker also allows us to extract the player strategy that achieves it. This process is known as strategy synthesis~\citep{kwiatkowska2016model2} and is used systematically throughout the CSG process in which we examine how strategies evolve over time as players adopt optimal strategies. Model checking is computationally expensive and so this approach would not be suitable for a more elaborate game, for example, where multiple objects have highly-precise 3D positions. Although PRISM has been used to verify soundness properties in simple 2D games~\citep{penguin}, most modern games are too complex to be modelled accurately in this way without overly compromising abstractions. For model checking to be feasible on current hardware, a system must have no more than roughly $10^{10}$ states.%a few billion states.  
\par
% There is an opportunity here to refer to research where chunks / subsystems of games have been model checked, like dialogue trees in Fallout, but I use that ref. all the time and it isn't very good.

In order to demonstrate that the outcomes of CSG represent a realistic evolution of a game, we required data from a real game that was elaborate enough to require considered decision making from players without being so complex as to prevent the use of model checking. We also required a system where player data could be compared for multiple configurations to allow a comparison of their respective balance. Finding a pre-existing game satisfying both of these requirements seemed unlikely, so we extended an existing case study to something more akin to a real game, with a view to developing it as a mobile application from which to collect data. This way we could perform CSG analysis (on models) to find candidate configurations that promised theoretical levels of balance, before releasing the game and testing how they performed in practice.
\par

% \vspace{0.5cm}
% \addlinespace

Having control of the game, and the data it allowed us to collect, put us in the unique position of being able to design a system to generate data that was not only useful to us (in terms of our game balance research), but also in its own right in the context of system modelling. It is our intention to publish our dataset in full in the near future. Real-world datasets of sufficiently well specified systems are rare, so we will specify our system and the nature of the data it produces in as much detail as possible --- however this is outside the scope of this paper. %due to understandable privacy issues relating to commercial products. \par
\par

\par
 
% =================================
\section{DESIGN DETAILS \& REQUIREMENTS}\label{app-des}

% It was necessary for RPGLite to deliver on some core requirements. It had to deliver a dataset that could accommodate two immediate use cases, and could potentially provide utility to other researchers with unknown research aims. In addition, it had to conform to a specific system specification: RPGLite had to implement the mechanics and design of the game of the same name, as developed previously for theoretical research purposes.
% \par

% To explain this context, we here describe RPGLite's game design as defined by its rules, mechanics and configuration. 
% \par

RPGLite, the game, is defined by its \emph{rules}, \emph{mechanics} and \emph{configuration}. We present these here. In later sections, ``RPGLite'' is used solely to refer to RPGLite, the application. 
\par

\subsection{Rules}
RPGLite is a two-player, turn-based game in which each player chooses a pair of unique characters from a pool of eight. Each character has a unique action and three attributes: health, accuracy and damage. Some have additional attributes described by their action. On their turn, a player chooses the action of one of their alive characters and targets an opposing character with their action. That action will succeed or fail based on the acting character's accuracy value. Players can choose to skip on their turn or to forfeit the game at any time. A coin is flipped to decide which player goes first and the winner is the player who is first to reduce both of their opposing characters health values to 0. 
\par

\subsection{Mechanics}
The mechanics of RPGLite are encapsulated in the eight characters and their actions:
\begin{description}
    \item[Knight:] targets a single opponent;
    \item[Archer:] targets up to two opponents;
    \item[Healer:] targets a single opponent and heals a damaged ally or themselves;
    \item[Rogue:] targets a single opponent and does additional damage to heavily damaged targets;
    \item[Wizard:] targets a single opponent and stuns them, preventing their action from being used on their subsequent turn;
    \item[Barbarian:] targets a single opponent and does additional damage when heavily damaged themselves;
    \item[Monk:] targets a single opponent and continues turn until a target is missed, and;
    \item[Gunner:] targets a single opponent, does some damage even on failed actions;
\end{description}
The additional attributes needed to describe the characters fully are the \textit{heal} value of the Healer, the heavily damaged value for the Rogue (the \textit{execute range}), the heavily damaged value for the Barbarian (the \textit{rage threshold}), the increased damage value for the Barbarian (the \textit{rage damage}) and the miss damage (\textit{graze}) for the Gunner. 
\par

\subsection{Configuration}
In total there are 29 attributes for the characters in RPGLite. A configuration for RPGLite is a set of values for each attribute. These attributes are the parameters we tune in an attempt to balance the game. The application was released with a configuration which we suspected of being balanced based on automated analysis. After a significant number of games were played, the application was updated with a new configuration (dubbed ``season two''), with the aim of maintaining player interest. The new configuration had altered attributes for seven of the characters, for example the Healer's health value decreased from 10 to 9 and their accuracy increased from 0.85 to 0.9. Only the Wizard remained the same between configurations.
\par

% =================================
\section{THE FINAL PRODUCT}\label{impl}

As context for the reflections in the rest of this paper, it is necessary to describe the application that was actually built. This finished product is a combination of implementation details and the design decisions that lead to their implementation.
\par

\subsection{Design}
RPGLite was designed to be simple to understand and play, so as to keep players interested and reduce barriers to entry. On logging in, players are presented with five ``slots'' for games, which can have a number of states:

\begin{itemize}
    \item Unused, waiting for a game to be made
    \item Added to a queue of players waiting for a random match to be made
    \item In an active game, \begin{itemize}
        \item Waiting for a move to be made by the player 
        \item Waiting for an opponent to make a move
    \end{itemize}
\end{itemize}

\begin{figure}[h]
    \centering
    \includegraphics[width=1.65in, trim= 0 58px 0 52px, clip]{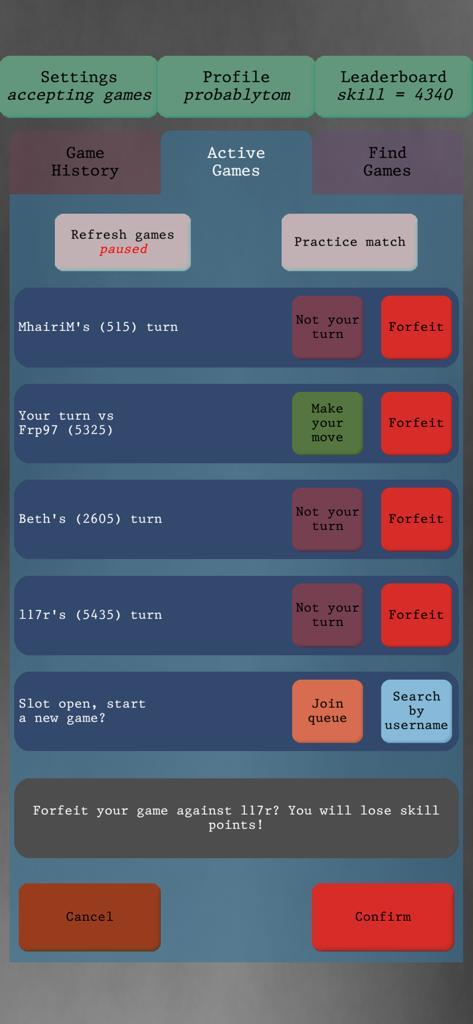}
    \includegraphics[width=1.65in, trim= 0 58px 0 52px, clip]{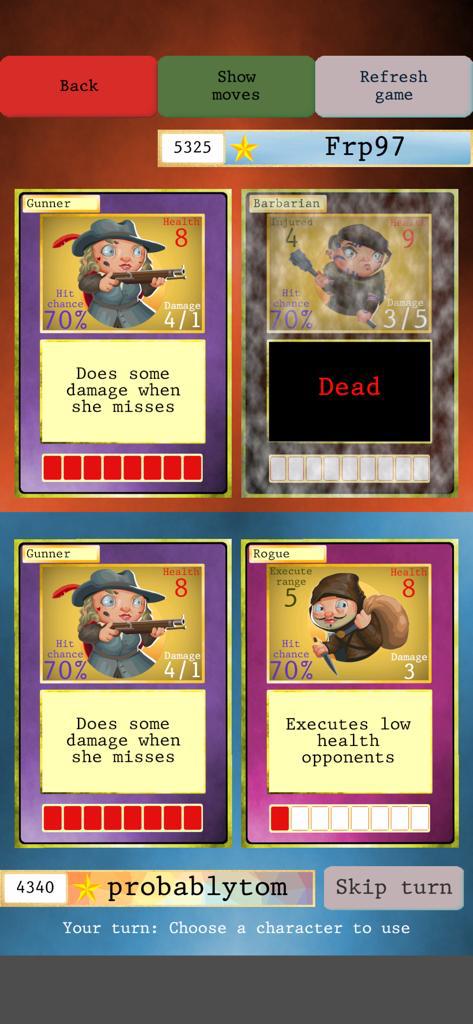}
    \caption{Screenshots of the released application showing a player's home screen (\textit{left}) and a game in progress (\textit{right})}
    \label{fig:central-screens}
\end{figure}

On starting a new game, players choose two unique characters from a set of eight and are presented with cards representing their chosen pair opposite against their opponent's. Animations of character cards are used to indicate whether the player can make a move, as well as an on-screen prompt. The application was designed to be as frictionless as possible to use, although, as discussed in lesson 4, we found some users were still confused and we simplified the design further via iteration.
\par

\begin{figure}[h]
    \centering
    \includegraphics[width=1.65in, trim= 0 0 0 79px, clip]{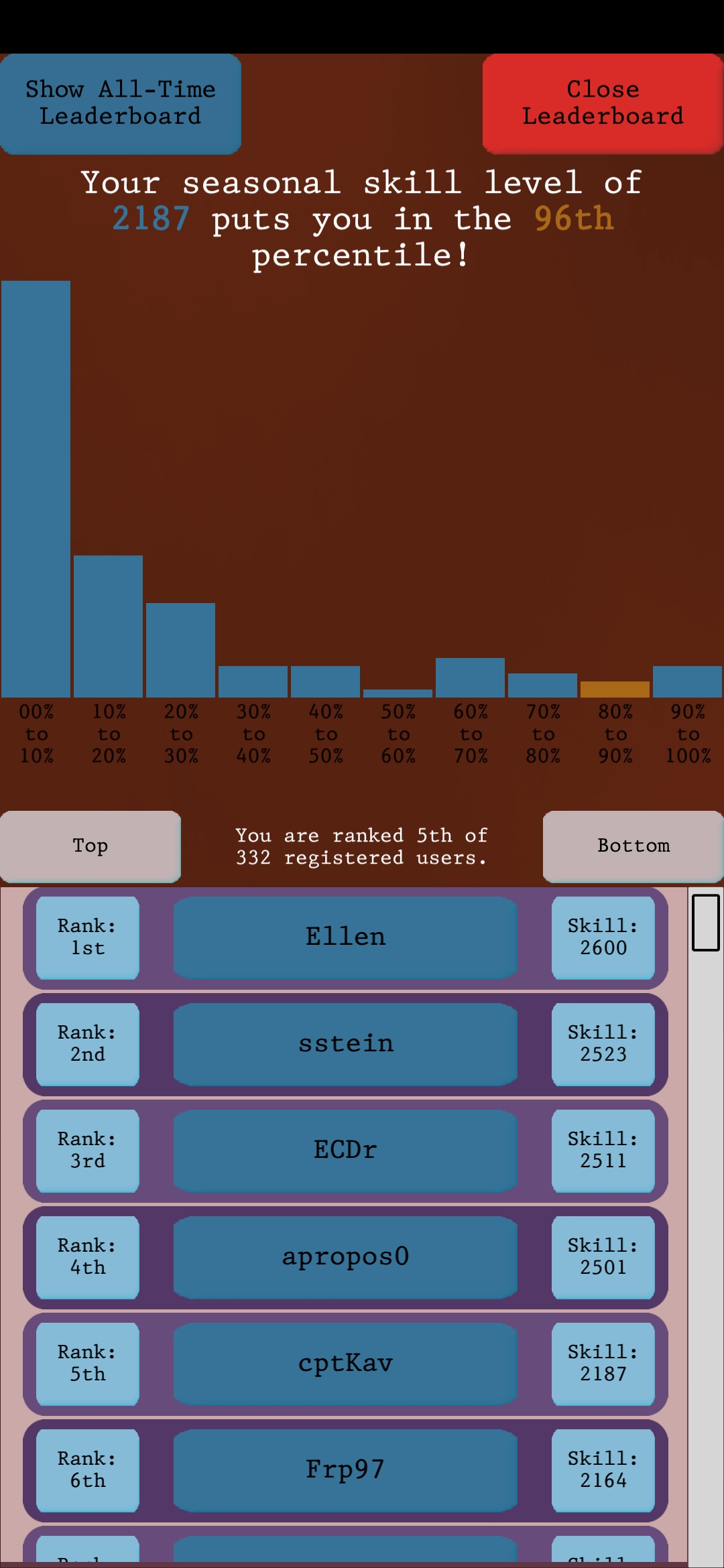}
    \includegraphics[width=1.65in, trim= 0 0 0 79px, clip]{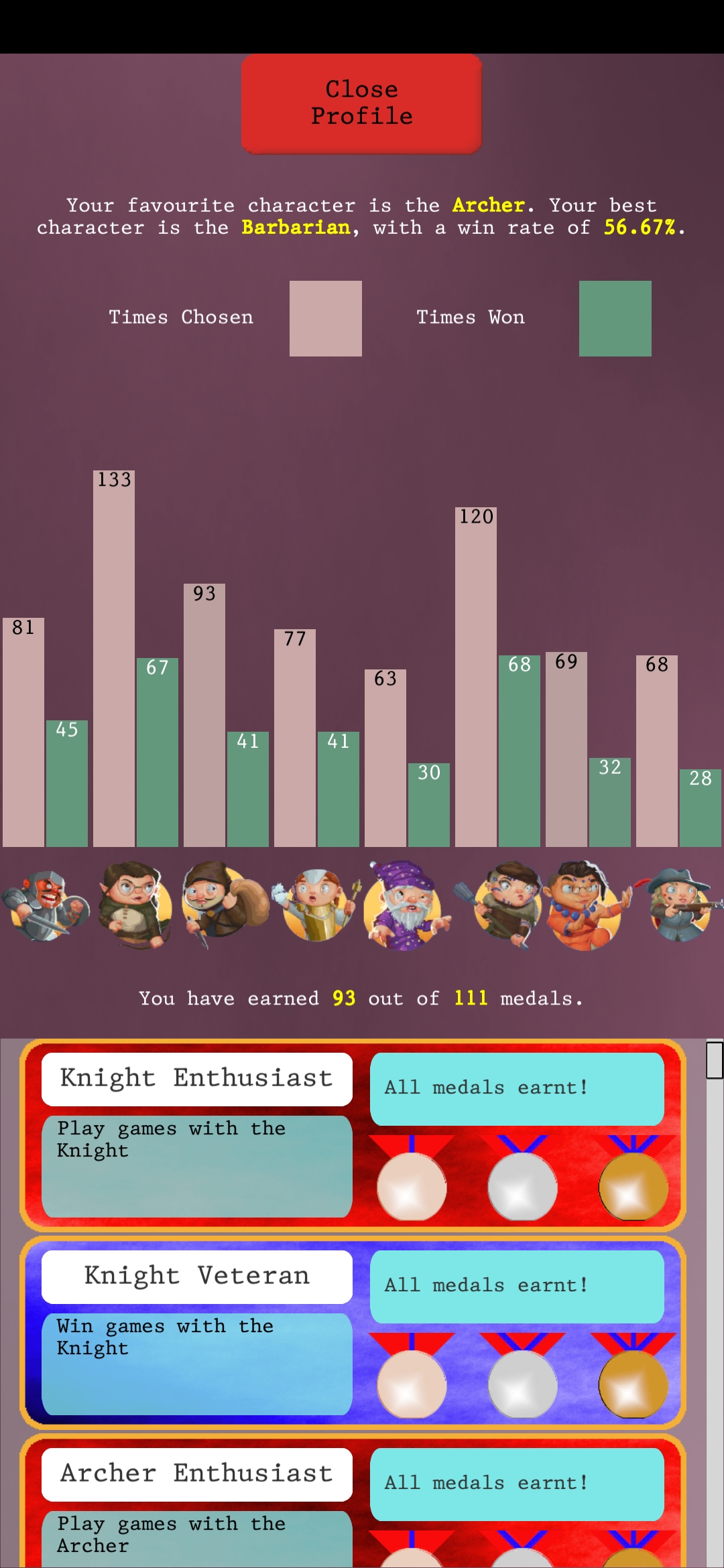}
    \caption{Screenshots of peripheral systems implemented towards the end of development. The leaderboard screen (\emph{left}) shows a player's skill points compared to all others. The player profile screen (\emph{right}) displays usage data for all characters and the medals earned.}
    \label{fig:peripheral-screens}
\end{figure}

Additional features were implemented specifically to encourage player retention. As can be seen in~\cref{fig:peripheral-screens}, peripheral systems around the core game, such as medals for players to earn and leaderboards to climb, were intended to give players goals to achieve and a reason to stay invested in the game.
\par

As players were technically experiment participants, it was necessary to have them ``sign'' an ethics-approved consent form and be delivered an information sheet. We implemented this by requiring players to scroll through a panel containing their consent form and information sheet on registration, and explicitly tick boxes confirming that they consented to all necessary parts and were over 15 years old.
% We experimented with more complex systems, including delivering to them consent forms signed (by their username) and dated electronically after successful registration, but found it was beyond our requirements.
\par

\subsection{Implementation}
RPGLite was implemented as a mobile game written in Unity. We chose Unity for it's ability to compile the same project to both iOS and Android and it's active community with numerous video tutorials for beginners. RPGLite made requests to a public-facing REST api, written in Python3 and run on university-provided servers with a firewall under the control of the institution's IT services. This server initially handed data processed in the client to the database to avoid a direct connection (and the risk of exposing the database publicly), but became a larger aspect of the engineering as discussed in lesson 3. The project collected and stored its data within a MongoDB database also hosted locally within the university.
\par

% T: I think we can do a better image for this section in design instead of impl...?
% \begin{figure}[h]
%     \centering
%     \includegraphics{images/RPGLite_network_architecure.pdf}
%     \caption{The structure of the final network architecture.}
%     \label{fig:network-architecture}
% \end{figure}

% =================================
\section{LESSON 1: RESIST TEMPTATION}\label{lesson:temptation}

% Here's an outline with a lovely example
At many points in the development process, we found it difficult to constrain the feature set of the end product. The unbounded nature of the project led to additional features being implemented as development became a ``labour of love''. These delayed the delivery of the game, and few new ideas were actually discarded. Only some of these features were beneficial to the player experience. An illustrative example is the comparison of two such ``peripheral systems'': the leaderboard and players' profiles. Players load the leaderboard roughly three times as often as their profiles and, anecdotally, they are a central component of player retention. An equal amount of effort was spent on each. During development it was impossible to know how often a feature would be used in practice. 
\par

% Here's a problem 
The ideas that came to us during development were sometimes essential to the project's success, and to resist all of these would have resulted in a poorer product. The danger we identified in our own endeavours was a desire to implement these ideas \emph{for their own sake}, and not for their benefit to our end product. New ideas must be abandoned where their benefit does not outweigh the additional time they would demand. An agile development approach is the best in these scenarios, where requirements naturally change over time. 
\par

% Here's a similar one
Much like implementing new features, we found that the refinement of existing features risked an emotional investment. Existing design components, such as colour schemes and layouts of minor UI elements, were constantly changed prior to release. We found the adage, ``don't let the perfect be the enemy of the good'', useful in such moments.
\par

% Here's why we struggled with them AND how we would fix it if we were to do it again.
We struggled to resist temptation because of our inexperience with app development, our lack of a thorough plan and the fact that we were co-developing and therefore reticent to shoot down each other's ideas. For other developers in similar situations to our own, we recommend a more structured approach. Firstly, a project should have a plan produced at its inception, which is maintained throughout the development process. Second, we suggest adding to this plan a ``margin''; a block of unallocated time at the end of the project that can be spent on developing new ideas. As development progresses, this margin can be ``spent'' on new ideas or refinements to existing design elements. This facilitates necessary discussions by framing them within the context of a shared resource.
\par

\section{LESSON 2: EMPLOY AVAILABLE RESEARCH NETWORKS}\label{lesson:research-networks}
% William

% For your consideration to include somewhere here
% While we are unable to identify individual players from collected data, we have a strong suspicion that dedicated players are typically those who we knew prior to play. This is certainly a weakness of our data. An ulterior motive of this paper is to announce the game within the academic community. We are confident that the dataset produced will be of interest to a broad range of researchers, and anticipate that, should we require further users, we would have to do so outside our personal networks, the majority of whom are now playing the game where possible. In this way we intend to take our own advice, and make use of an active game design and research community who might be interested in the dataset we intend to release.\footnote{For interested persons, we nominally intend to collect data via this game until April 2021. The game can be downloaded via \href{https://rpglite.app}{https://rpglite.app}.}

% Outline with example.
% Problems (1-n)
% Solution

Advertising is a major cost of app development; new users are expensive. With no money for player recruitment we were forced to promote the application in a similar way to other research experiments within a university context, through participant calls in mailing lists and departmental announcements. Beyond this we sought out opportunities for free publicity from within our research community. We found that there is an appetite for open data and by encouraging people to play our game ``for science'' our promotions were better received. We anticipated undergraduate students would make up the majority of our users. However, while promotions targeted at undergraduates introduced a large number of users, those users tended to only complete a few games before stopping. For our research we wanted to investigate how players learn over time, we needed high player retention to allow users time to ``learn'' the system. We observed that retention was highest within players who had a vested interest in us or the research itself, or when the game was adopted by users from a social clique. 

\begin{figure}[h]
    \centering
    \includegraphics[width=3.3in]{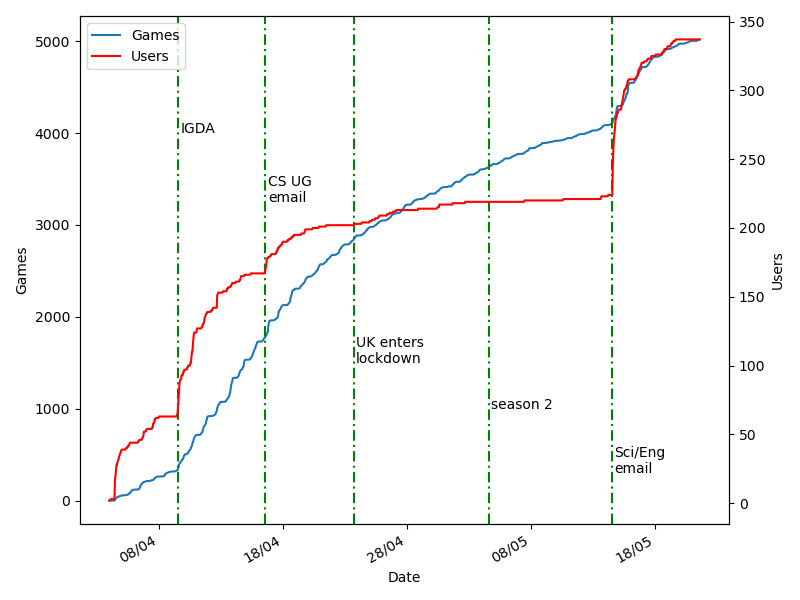}
    \caption{The rate of user acquisition in the weeks following RPGLite's release. Important events are also marked: promotion of the application through the Scottish International Game Developers Association branch, an email to Computing Science undergraduates, the date from which UK citizens were told to stay inside if at all possible, the time of a major update to the game and an email to all Science and Engineering undergraduates at the University of Glasgow.} % could be rephrased.
    \label{fig:acquisition}
\end{figure}

\begin{figure}[h]
    \centering
    \includegraphics[width=3.3in]{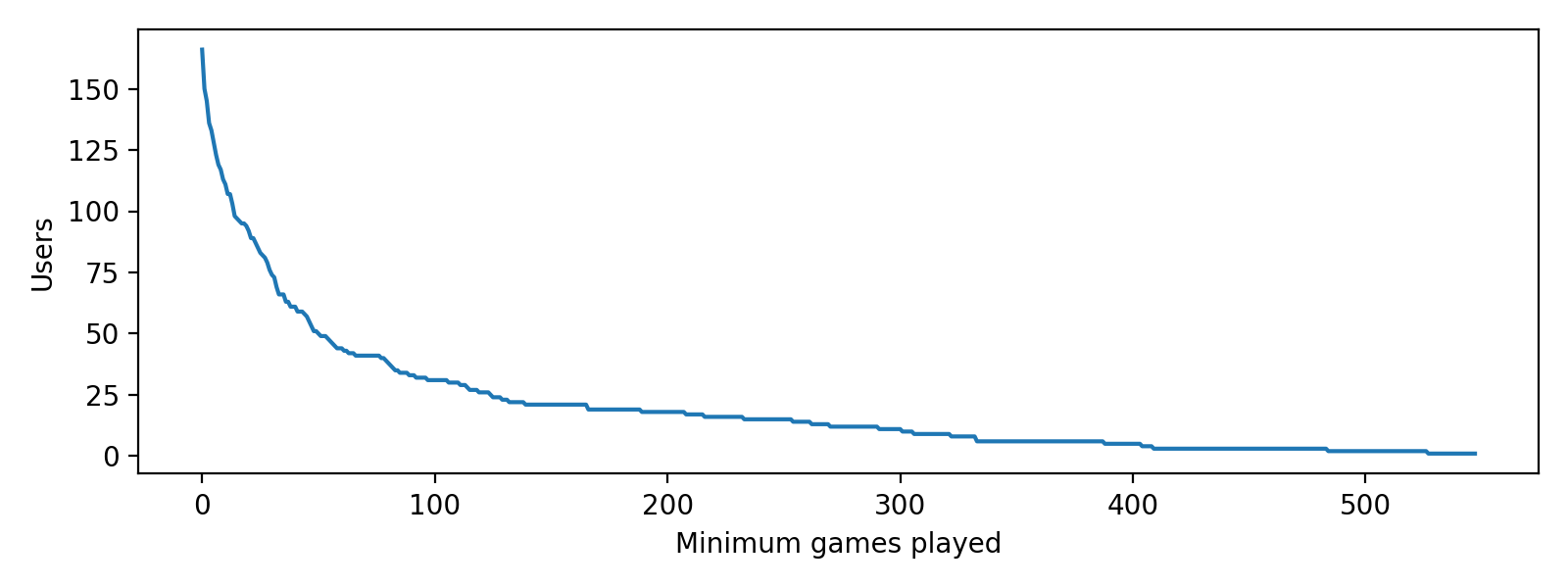}
    \caption{The number of users to have played at least a given number of games.}
    \label{fig:retention}
\end{figure}

In comparing events that we expected would increase player numbers with their effects on new users and games played (a suitable measure of data generation)~\cref{fig:acquisition}, the retention of the different groups recruited is pronounced. Over half of our users failed to successfully complete a single game, and several users installed the app without registering an account. We are fortunate enough to know the chair of the Scottish branch of the International Game Developers Association (IGDA) who kindly shared an advert for the game. The increase in the speed of game completions accompanying the influx of new users from his involvement shows that those players were valuable data generators. The figure also shows that the large intake of undergraduate students from Science and Engineering only caused a brief uptake in activity, which quickly dissipated. We believe this is due to either the lack of a relationship with us as the developers or of interest in games research. We also assumed that a large update might increase activity, but found that not to be the case. A single large update changing the configuration of the game, adding seasonal leaderboards and improving existing features had no noticeable effect on the number of games completed. The extent to which our data comes from a small subset of users is shown in~\cref{fig:retention}.

Throughout development we sought advice from those around us with relevant experience. Many of our university colleagues had been involved in various aspects of application development and deployment, and advised us throughout. For example, a web designer gave advice on UX design and a gamification researcher suggested various incentivisation systems. We also relied heavily on our department's IT services team for support in deploying the middleware server and administrative staff for promoting the app once it had been released. Application development is multifaceted and the support of our peers was important in areas where our skills were insufficient.
\par

% Summing up.
% WILLIAM NEW
Without the extensive use of the research communities we belong to, RPGLite would have been an inferior application, producing a less rich dataset. There are numerous skills required to develop a system that people will use willingly. Engaging peers early in the process and being clear in your aims will highlight the areas in which you need support. Where user retention is important your research community is vital, as they already have a connection to you which will see them invested in the project from the outset. Your individual network is unlikely to be enough to generate a significant dataset, so we recommend engaging colleagues to advertise on your behalf. RPGLite never sought to compete with professionally developed games, but through our various communities we manged to generate enough interest for a steady playerbase.

% OLD -- seen as 'too weak a conclusion to L2'
% We encourage developers in similar situations to engage their colleagues where possible. Their experiences of developing and deploying similar applications will be beneficial. 
\par

%Our research communities gave salient advice throughout the project and were invaluable in player retention and acquisition. As resources for player recruitment were non-existent, the support of our fellow researchers was vital.

% =================================
\section{LESSON 3: THE SMALLER THE CLIENT, THE BETTER}\label{lesson:thin-client}

% Problem description.
The one aspect of RPGLite's implementation that we most regret is the amount of game logic being delivered to players in the mobile client rather than the server. There are many reasons for this, the main one being that the server could be replaced immediately if a bug were to be found. This is in contrast to compiling, re-installing and re-testing attempts to fix the given bug, were it to reside in the client. Fixing server-side bugs allowed more rapid iteration when fixing those with origins we did not understand.
\par

% A pertinent example.
% Conversely, bugs in production code in the client were very hard to quickly fix. Getting the app through the app stores could be troublesome.
The need for moving logic out of our client became apparent after we had pushed production code to app stores and had real players taking part in our experiment. A particularly dedicated player discovered a bug where, after playing enough games, characters that had been unlocked through repeated play would become locked again and could no longer be accessed. Had this bug been in the server, the issue could have been fixed, and a new version deployed in seconds that lightweight clients could connect to. With our larger client, this required testing in Unity, testing on-device (to ensure that there weren't platform-specific bugs), and deployment to app stores for approval and distribution. This process took days, even though the bug was trivial to fix.
\par

Large clients also risk introducing a duplication of code when paired with a secure server. To validate game logic computed by a client, servers must replicate much of the processing the client previously performed, to verify that a malicious user hasn't supplied corrupted game states. This process requires the implementation of game logic within the server. As a result, a secure server must include game logic regardless of whether the client does. This means spending time, an already scarce resource, on duplicated code. This is another reason we recommend developing a lightweight client, leaving the majority of computation to a larger server.
\par

% Times we thinified the client and the sections that we moved

When we realised that we had produced a large client, we made efforts to move to a more server-centric design. For example, we considered sending push notifications via APIs directly written into our client. However, the flexibility and control of implementing this server-side caused us to move our notification code to the server. After this, we implemented much of our peripheral systems logic in the server, including the leaderboard, medal logic, password reset, and much of the matchmaking systems.
\par

Overall, we found that areas where the client was lightweight allowed more rapid prototyping and bugfixing. We recommend other projects be constructed with a small client for these reasons, as well as avoiding duplication of code and a reduction in application size by limiting client-side dependencies.
\par

% for potential summary paragraphs
% It is notable that these issues would have been less apparent under rigorous testing, which we did not employ.
% =================================
\section{LESSON 4: TEST EARLY, TEST OFTEN}\label{lesson:testing}
% William

% Testing good

The best source of feedback and advice we received was from the shared document we circulated alongside our two private test releases. We specifically chose friends and colleagues who knew us well enough to be able to have honest discussions on the weaker aspects of the application. We carried out the testing by sharing Android application packages with Android users and inviting iOS users to participate in private beta testing via Apple's TestFlight system. We were able to implement the majority of the suggestions made, many of which have become central components in the final game. This stage highlighted the importance of push notifications and streamlining the user experience. Specifically, our test users found that they would often forget to check whether they had moves to make. Before testing we had investigated the feasibility of implementing push notifications, but were unsure if they were worth the time to develop. Following testing feedback, we made this a priority.
\par

% clears examples of why testing good.
% We also found that almost all of our test users complained that they were prompted to view a tutorial too often and they didn't like waiting through long roll animations. We learned, via frequent testing, that quickly getting into and out of games was important for users, and that as a result we ought to bother players as infrequently as possible. Without a focus on real-world testing and frequent iteration, this would have been very difficult to discover.
% \par

\begin{figure}[h]
    \centering
    \includegraphics[width=1.1in, align=c]{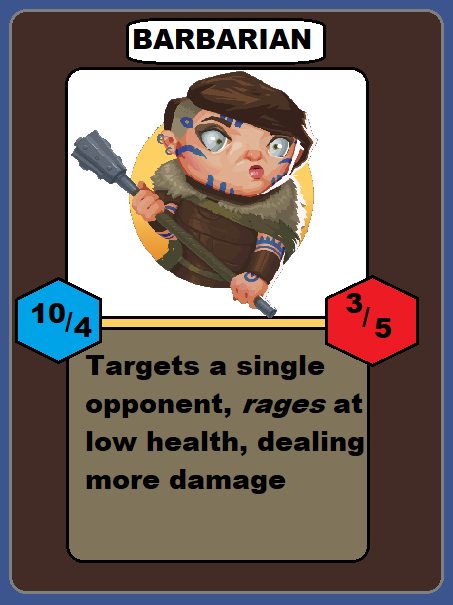}
    \includegraphics[width=1.05in, align=c]{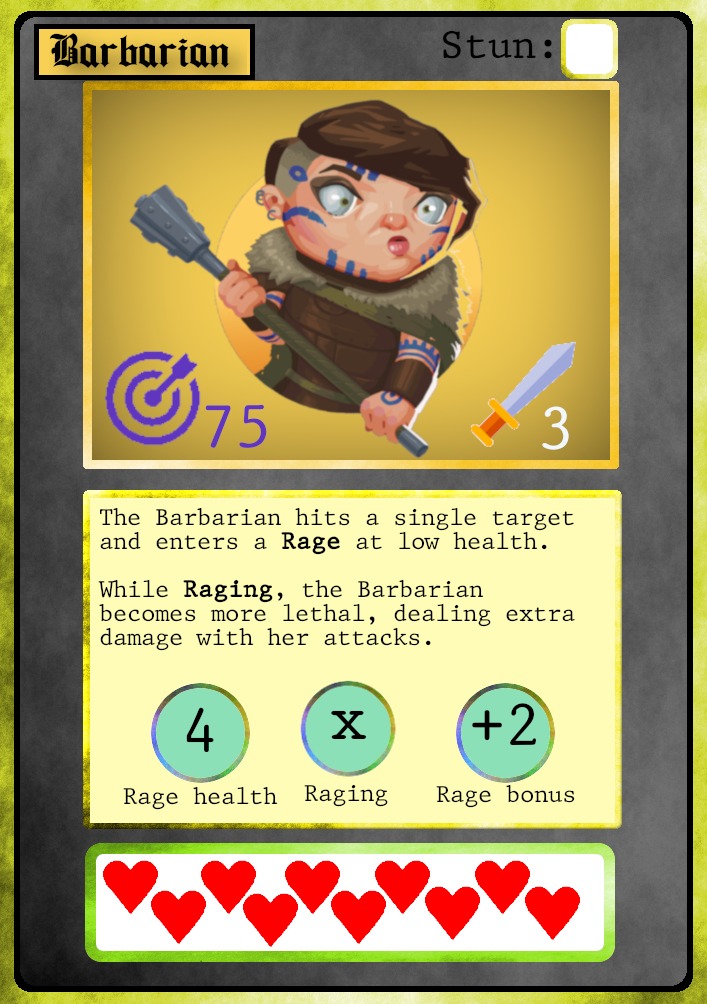}
    \includegraphics[width=1.05in, align=c]{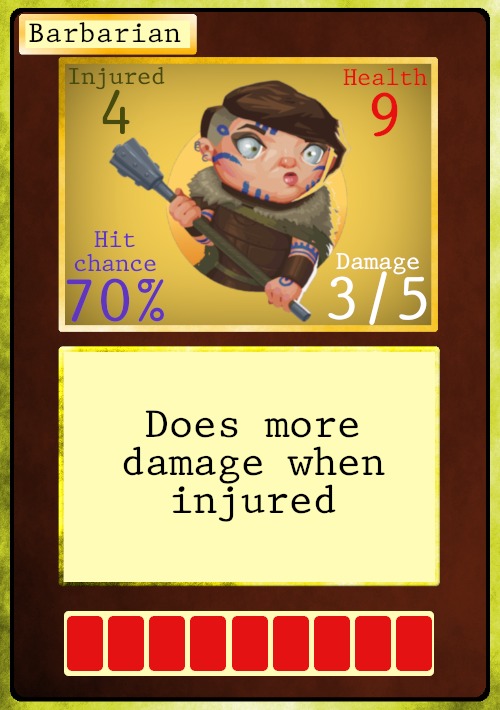}
    \caption{Evolution of the Barbarian card artwork throughout the design process from initial prototype (\textit{left}), to internal testing version (\textit{centre}) and current version (\textit{right})} % could be rephrased.   T: I quite like it! Think this is fine
    \label{fig:card-compare}
\end{figure}

% ==== BEGIN TOM'S NONSENSE
% T: I know this is longer than you wanted --- will let you trim as you see fit
% Real-world testing was tricky for certain components of the game --- particularly via Apple's TestFlight system. This was due to a number of compounding issues. Certificates for RPGLite were distributed via our institution's apple developer account, and our control over this was limited. This meant that beta testers' devices could not be added to an approved list for internal testing, forcing beta builds to undergo a lengthy validation process for public beta approval. Some features --- notably push notifications --- would not work in simulators. The combination of lacking simulator support and hours to days of delay to test on a physical device meant that some iOS-specific bugs persist in production code, despite attempts to fix these.
% ==== CEASE TOM'S NONSENSE

% specific examples of test feedback that was implemented
The user interface, colour scheme and card art of the final application are a result of feedback from our test users. As shown in \cref{fig:card-compare}, the cards went through a series of designs. Responding to test feedback that character cards were too complicated, the final designs were significantly simpler. We also received specific advice, such as blacking out the action description of a stunned character to make it clear that they could not act. Having an ongoing dialogue throughout development with invested parties, meant that we could rapidly pivot to accommodate their suggestions.
\par

% practice analysis on test data to flesh out final dataset.
From analysis of our test data we discovered a gap between the data we were collecting and possible useful information we could capture. Specifically, we realised we could log user interactions with the application, noting actions they performed, when they performed them and what the result was if any (for example, ``a user searched for another by their username and found they had no free game slots''). This idea was a result of realising that even amongst our dozen test users, there were distinct styles of interacting with the application. We thought that classifying these interaction styles would be of interest.
\par

% Lesson summary
Testing allowed us to identify areas in both the application and the dataset that were lacking. We would encourage future researchers to get early versions of their applications into the hands of testers multiple times before finalising their system. There were many improvements made to RPGLite specifically because we had others test it, and could assess it across a suite of target devices. We structured the format of the feedback we received from testers in our shared document by grouping requested feedback under specific headings and directing them to features in which we lacked confidence. This helped to scaffold the insightful conversations amongst our test users, and we strongly recommend others make an effort to facilitate a similar dialogue.
\par

\section{CONCLUSION}\label{conclusion}
In releasing RPGLite we learned several lessons about the realities of mobile game development within research. We have outlined our key insights and hope that these will be helpful to researchers developing similar tools. To summarise, the lessons that we learned are: to beware of scope creep and lengthy feature refinement
%and <etc>
; to utilise ones research community for their expertise and willingness to contribute; to structure the application to permit rapid bug-fixing, and to avoid duplication of code, and; to test as soon as you have a workable build and to continue testing up until release.
\par

Pausing our research to develop a mobile game was an atypical activity. We hope that these observations are helpful to other researchers developing similar projects. If they are, we encourage them to document the methodologies they follow for building data-generating games for the benefit of others engaged in similar projects, and the lessons they learned doing so.
\par

% Moving to future work
% We intend to release the full dataset gathered alongside the client and server code and our own research that motivated its development, in the near future. 
\par

% =================================
\section{FUTURE WORK}\label{future_work}
% T: Turns out this is a required section from EUROSIS' format
% T, later: or is it?

% One solution to this problem is to generate plausibly realistic data about real-world systems, so as to sup- plement existing datasets with high quality synthetic ones. Thus alongside the collection of real data from our system, we developed parallel synthetic datasets for comparison. One of our future goals is to measure the similarity between these datasets and the collected data.
% Given a gap in the literature pertaining to sufficiently high-quality and well-specified datasets, two contri- butions must be made. The first, which we intend to provide in a forthcoming paper, is to fill this gap by releasing such a dataset. The data we intend to release contains details of play style, alongside individuals’ in- teractions with the application itself, so as to increase its relevance to the wider research community. We an- ticipate that this data will therefore be of interest to researchers in a variety of areas.
% The second is to address the lack of projects focus- ing on similar data collection. To address this, we
% here describe the lessons learned in producing and distributing a medium-scale online game for research purposes. Our aim is to reduce the barrier to entry for researchers embarking on similar projects in the future.

This paper details the experience of developing a mobile game for data collection. The next stage of our research is the processing and analysis of this dataset. We intend to explore many research questions using it, with some pertaining to the dataset itself and analysis of optimal play, and others, to the accurate simulation of RPGLite players.
\par

% The future work produced by RPGLite falls under three categories: Work pertaining to RPGLite itself, and work in the disparate research areas of the game's development team: system simulation.% and game modelling.

We will release the full dataset collected by RPGLite alongside the code constituting the game client and server in a future publication. This will include collections of all players, all games played, and all interactions recorded within the application. 
In addition, this dataset will include complete information about the games played such as moves made, characters chosen, and other details used in our own research. 
These collections include all the attributes we envisaged as being useful to future research. %; for example, how often a player chose each character, all moves made and players engaged in completed games, and more.
For example, a player document includes their username, played/won counts for each character, other players they have lost games against, skill points, and more.
We intend to omit only sensitive details, as all collected data is anonymised, and users have indicated through our registration process their consent for collected data to be disseminated through the academic community in the spirit of open science.
\par

\subsection{System Simulation}
Datasets sourced from sufficiently scaled and well-detailed systems are rare. Some are made available for academic use~\citep{bpi2015dataset}, but available data typically originates from large industrial systems lacking public specification for competitive reasons, or from well-scaled systems which lack the supporting detail to be useful. We are therefore interested in taking small datasets from systems of a manageable size, and producing supplementary synthetic data which appears plausibly realistic. We believe this data can be produced by an application of aspect-oriented programming~\citep{wallis2018modelling}. A small, naive simulation of behaviour is modified via applied aspects to introduce errors and improvements. We are in the process of developing simulations of RPGLite play and aspects to improve the simulation's realism. We aim to verify that this produces plausibly realistic synthetic datasets by comparison with RPGLite's empirically sourced data.
\par

Assuming this work is successful, we intend to show that aspects can ``fit'' themselves to real-world data. We expect these to produce datasets with optimal similarity to empirical counterparts via the application of genetic algorithms on their parameters\citep{wallis2018process}. A corollary of this approach would be that, in addition to highly realistic simulations, aspect parameters would then describe the nature of real-world agents. This process could then be used as a lens through which to analyse actual behaviour, weighing various influences by their importance.
\par

% Assuming this work is successful, we intend to show that aspects can ``fit'' themselves to real-world data by repeated modification of their parameters. We expect that this genetic algorithm will discover parameter combinations which produce datasets with optimal similarity to an empirical counterpart~\citep{wallis2018process}. 

% These parameters might describe the nature of the agents being modelled in the real world. This process could then be used as a lens through which to analyse actual behaviour, weighting various influences by their importance.

% We are also eager to perceive whether these parameters might describe the nature of the agents being modelled in the real world, by indicating the aspects which model a real influence on behaviour via their degree of application. 
% \par

\subsection{Game Development and Player Analysis}

% Game development
As described in the motivation, we have developed tools that use model checking to assess game balance without gameplay data. RPGLite was originally intended solely to verify this process with both quantitative analysis and qualitative player feedback. Based on the findings of our model checking analysis, we believe both of the configurations released for RPGLite are balanced, but one is ``more balanced'' than the other. Calculating the extent of this and comparing our metagame predictions to what was observed in the dataset when players explored RPGLite will measure the validity of our approach.

% Player analysis
RPGLite is a bounded system that can be model checked, this allows for highly specific analysis of player actions. We can calculate the cost of any move made in the game as the difference between the player's subsequent probability of winning and their probability having chosen the best move available. By comparing the costs of the moves a player makes over time we can measure their rate of learning without considering their opponents. The effect of having definitive measures of player mistakes for gameplay analysis is a research area which is of great interest to us. 
% WILLIAM NEW
This could help us answers questions about the situations in which players make mistakes and what causes them. Beyond game research, this could potentially lead to aiding the design of systems which aim to minimise human interaction errors.

% =================================
\section{ACKNOWLEDGEMENTS}\label{acknowledgements}
We could not have released RPGLite without significant input from our colleagues and friends. In particular, we would like to acknowledge Ellen Wallace, Marta Araldo, Justin Nichol, Craig Reilly, Adam Elsbury, Alistair Morrison, Chris McGlashan, Frances Cooper, Brian McKenna, and our test players. The work was partly supported by Obashi Technologies.

% =================================
\bibliography{refs}

\end{document}